\documentclass[aps,prd,twocolumn,showpacs,floatfix,preprintnumbers,amsfont,amsmath,amssymb,nofootinbib, superscriptaddress]{revtex4-1}

\usepackage{hyperref}
\usepackage{tabularx}
\usepackage{ctable}
\usepackage{graphicx}
\usepackage{dcolumn}
\usepackage{bm}
\usepackage{siunitx}
\usepackage{mathtools}

\usepackage{color}

\usepackage{array}

\newcommand{\PreserveBackslash}[1]{\let\temp=\\#1\let\\=\temp}
\newcolumntype{C}[1]{>{\PreserveBackslash\centering}p{#1}}
\newcolumntype{R}[1]{>{\PreserveBackslash\raggedleft}p{#1}}
\newcolumntype{L}[1]{>{\PreserveBackslash\raggedright}p{#1}}

\def\rmd{{\rm d}}
\def\bfchi{\boldsymbol{\chi}}

\newcommand{\be}{\begin{equation}}
\newcommand{\ee}{\end{equation}}
\newcommand{\ba}{\begin{eqnarray}}
\newcommand{\ea}{\end{eqnarray}}

\newcommand{\reffig}[1]{Figure~\ref{fig:#1}}

\newcommand{\reftab}[1]{Table~\ref{tab:#1}}

\def\CANDIDATE{GW151216 }

\begin{document}

\preprint{APS/123-QED}

\title{A Highly Spinning and Aligned Binary Black Hole Merger in the Advanced LIGO First Observing Run}

\author{Barak Zackay}
\email{bzackay@ias.edu}
 \affiliation{\mbox{School of Natural Sciences, Institute for Advanced Study, 1 Einstein Drive, Princeton, New Jersey 08540, USA}}
 \author{Tejaswi Venumadhav}
\affiliation{\mbox{School of Natural Sciences, Institute for Advanced Study, 1 Einstein Drive, Princeton, New Jersey 08540, USA}}
\author{Liang Dai}%
\affiliation{\mbox{School of Natural Sciences, Institute for Advanced Study, 1 Einstein Drive, Princeton, New Jersey 08540, USA}}
\author{Javier Roulet}
\affiliation{\mbox{Department of Physics, Princeton University, Princeton, New Jersey 08540, USA}}
\author{Matias Zaldarriaga}
\affiliation{\mbox{School of Natural Sciences, Institute for Advanced Study, 1 Einstein Drive, Princeton, New Jersey 08540, USA}}

\date{\today}
             
\begin{abstract}
We report a new binary black hole merger in the publicly available LIGO First Observing Run (O1) data release. 
The event has a false alarm rate of one per six years in the detector-frame chirp-mass range $\mathcal{M}^{\rm det} \in [20,40]M_\odot$ in a new independent analysis pipeline that we developed. 
Our best estimate of the probability that the event is of astrophysical origin is $P_{\rm astro} \sim 0.71\, .$
The estimated physical parameters of the event indicate that it is the merger of two massive black holes, $\mathcal{M}^{\rm det} = 31^{+2}_{-3}\,M_\odot$ with an effective spin parameter, $\chi_{\rm eff} = 0.81^{+0.15}_{-0.21}$, making this the most highly spinning merger reported to date. 
It is also among the two highest redshift mergers observed so far. 
The high aligned spin of the merger supports the hypothesis that merging binary black holes can be created by binary stellar evolution.
\end{abstract}

\maketitle


\section{\label{sec:Intro}Introduction}

The LIGO/Virgo collaboration has reported ten binary black hole (BBH) coalescence events detected during their First and Second Observing Runs~\cite{LIGOScientific:2018mvr}. 
These systems consist of black holes (BHs) with masses ranging from $10\,M_\odot$ to $60\,M_\odot$, with the primary and the secondary BH having comparable masses. 
Two of the detections, GW151226~\cite{abbott2016gw151226} and GW170729~\cite{LIGOScientific:2018jsj}, show conclusive evidence for at least one component BH having a positive spin along the direction of the orbital angular momentum, while the remaining events are consistent with both components being non-spinning. 
The astrophysical formation of the LIGO/Virgo BBHs is currently an active topic of research.


The detection of BBH signals is currently limited by confusion with noise transients of non-astrophysical origin. This sets the threshold matched filtering score for BBH triggers that can be confidently declared. We developed a new search pipeline~\cite{pipelinepaper} for which we made efforts to precisely characterize noise systematics and effectively mitigate noise transients~\cite{psddriftpaper, vetopaper}, with a view to reducing the detection threshold to search for faint and distant BBH events. Any addition to the BBH sample will bring considerable scientific value as it will enhance our ability to map out the BBH parameter space and accumulate evidence for or against models of their formation. In this paper, we report a new BBH merger event found in the publicly available LIGO First Observing Run (O1) data \cite{GWOSC} using the new search pipeline.
The strain signal recorded by advanced LIGO at the time of the event is consistent with the merger of two aligned and fast spinning BHs.

\section{The new event}

\begin{figure}[t]
    \centering
    \includegraphics[width=\columnwidth]{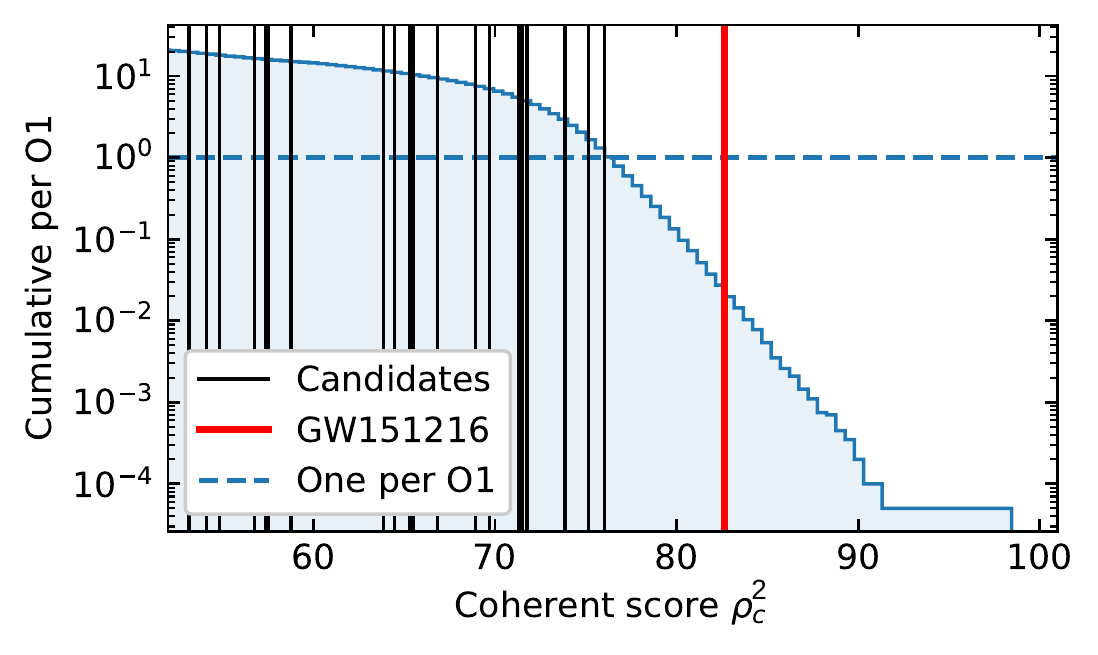}
    \caption{\label{fig:farplot} The blue curve shows the cumulative number of expected background events above a given value of the coherent score $\rho_c^2$ per O1 run in the \texttt{BBH 3} bank, estimated from \num{20000} timeslides of the data. The flattening at low values is an artifact of the threshold used while collecting background triggers. Vertical black lines mark candidates, i.e., triggers at physical shifts (with previously reported events and injections removed). The event \CANDIDATE, marked in red, has a FAR of 1 in 52 O1.}
\end{figure}
Ref.~\cite{pipelinepaper} reports the overall results of the BBH search we performed using our new compact binary coalescence detection pipeline. Our search used five banks, each covering a factor of two in detector-frame chirp-mass $\mathcal{M}^{\rm det} \equiv (1+z)\,(m_1\,m_2)^{3/5}/(m_1 + m_2)^{1/5}$. 

We detected a significant trigger within our template bank \texttt{BBH 3}, which covers the chirp-mass range $[20,40]\,M_\odot$~\cite{templatebankpaper}. Restricting the frequency range to $[20,\,512]\,$Hz in the analysis, we obtained a maximal network matched-filter signal-to-noise-ratio ${\rm SNR} = 8.5$ with the spin-aligned BBH waveform model \texttt{IMRPhenomD}. This trigger was previously reported as a subthreshold candidate in the 1-OGC catalogue by Ref.~\cite{NitzCatalog}, in which it was not deemed sufficiently significant to be declarable.  

This is the only significant new trigger we found in our analysis. The quality cuts we currently apply to the data do not yet reject all non-Gaussian noise artifacts (``glitches'') in the data, especially in the banks covering the heaviest masses. Improvements to our glitch-rejection algorithms may result in additional interesting triggers.

\begin{figure*}[t]
    \centering
    \includegraphics[width=\textwidth]{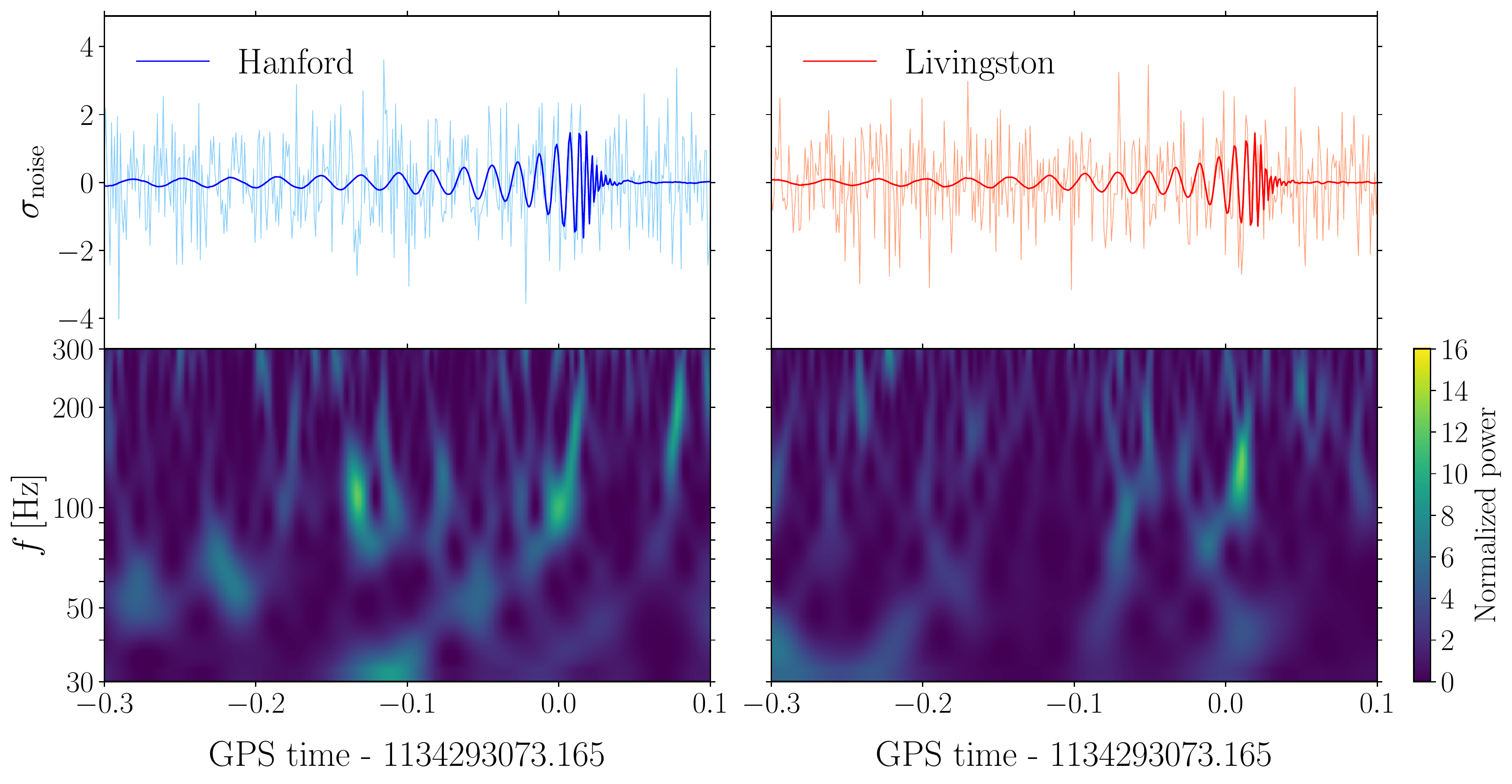}
    \caption{\label{fig:whitened_strain} Upper panels show the whitened strains around the trigger time of \CANDIDATE in LIGO Hanford/Livingston detectors (light colored curves). Overplotted are the maximum likelihood fits using the spin-aligned \texttt{IMRPhenomD} waveforms (dark colored curves). Lower panels show the corresponding spectrograms. Note that the best-fit gravitational waveform accumulates nearly the entire signal-to-noise in the frequency range $[30,\,300]\,$Hz.}
\end{figure*}

Figure \ref{fig:farplot} shows the background distribution of the coherent ranking score for coincident Hanford (H1) and Livingston (L1) triggers between $20 M_\odot < \mathcal{M}^{\rm det} < 40 M_\odot$, calculated empirically using $2\times 10^4$ different time slides of the data in the two detectors. 
The event has a false alarm rate (FAR) of 1 per 52 observing runs in the \texttt{BBH 3} template bank. 
The coincident on-time used for the analysis was 46.6 days, which implies that the inverse false alarm rate for the reported event in this bank is ${\rm FAR}^{-1}_{[20,40]} = 6.6 \, {\rm yr}$. This FAR, even after dividing by the number of binary black hole banks in our search is comparable to the one of GW170729 as reported in the GWTC-1 catalog \cite{LIGOScientific:2018mvr}.
Our best estimate for the probability that this event is of astrophysical origin yields $P_{\rm astro} \approx 0.7$, which is above the bar of $0.5$ for listing events in the catalog of confident gravitational-wave detections defined in \cite{LIGOScientific:2018mvr}.
For the computation of $P_{\rm astro}$ we used a bank rate measurement of one confident merger (with network SNR bigger than 100) per 30 days of coincident time. This estimate is based on the fact that six mergers appearing in GWTC-1 are in the relevant bank range, which gives a fairly well measured astrophysical rate.
We therefore adopt the name \CANDIDATE for this event.

No special artifacts are present at the time of the event or in its immediate surroundings.
Figure 2 presents the whitened strain data, best-fit waveforms, and spectrograms of the whitened data in the Hanford (H1) and Livingston (L1) detectors near the event. \reftab{parameters} presents the parameters of the event estimated using two different astrophysical priors, which, respectively, assume isotropic distributions for the individual spins of the component black holes, and a uniform distribution for the effective spin parameter that controls the waveform. In the two cases the effective spin of the event is larger than that of any gravitational wave merger observed to date; remarkably, the measured spin is close to unity under the uniform effective spin prior.


\section{Parameter Estimation}


We preprocessed the raw strain data as described in Sections C and D of our pipeline paper~\cite{pipelinepaper}. 
We estimated the PSD with Welch's method with a chunksize of \SI{64}{\second}, and used \SI{4096}{\second} of data to achieve adequate precision.
We downsampled the data to $\SI{1024}{\hertz}$, applied a high-pass filter to keep frequencies above \SI{15}{\hertz}, automatically flagged and removed loud non-astrophysical noise transients, and subsequently inpainted the masked data segments. 
Finally, we applied a a PSD drift correction to account for the non-stationarity of the Gaussian noise.

For the final analysis, we use $\SI{128}{\second}$ of Hanford/Livingston data around the trigger time for matched-filtering with template waveforms. 
We restrict to the frequency range $[20,\,512]\,$Hz, which is sufficient to analyze the coalescence of heavy binary black holes. 
We use the relative binning method to evaluate the likelihood of the data~\cite{Zackay:2018qdy, Dai:2018dca}, and couple it to the python package \texttt{pyMultiNest}~\cite{Buchner:2014nha} to generate samples from the posterior.

{\bf Spin-aligned model}. We first examine the source properties under the assumption that the spins of the two black holes are aligned with the orbital angular momentum. 
We fit the strain data using the phenomenological non-precessing waveform model \texttt{IMRPhenomD} for binary black holes~\cite{Khan:2015jqa}.
The intrinsic parameters that are varied are the component masses, $m_1$ and $m_2$, and the dimensionless aligned spins, $\chi_{1z}$ and $\chi_{2z}$, for the primary and the secondary black holes respectively. 
The model has seven extrinsic parameters to fully account for the correlations between the phase, amplitude and arrival time at the two LIGO detectors: the orbital inclination $\iota$, orbital phase $\varphi$, source sky location $({\rm RA},\,{\rm DEC})$, orbital roll angle $\psi$ on the sky, luminosity distance $D_L$, and the geocentric arrival time $t_c$.

The choices for the prior distributions of the extrinsic parameters naturally follow from the assumptions that the source is randomly located on the sky, and that the binary's orbital orientation is isotropically distributed. 
The merger rate as a function of the redshift is not known, so at the level of parameter estimation, we assume a prior distribution $P(D_L) \propto D^2_L$ up to $10\,{\rm Gpc}$ for the luminosity distance $D_L$, which corresponds to a constant merger rate per unit volume in Euclidean space.

As for the intrinsic parameters, we assign uniform priors to the component masses $m_1$ and $m_2$ (in the detector frame) within $[2,\,250]\,M_\odot$, while restricting the detector-frame chirp mass $\mathcal{M}^{\rm det}$ and the mass ratio $q = m_2/m_1$ to the ranges $\mathcal{M}^{\rm det} \in [10,\,40]\,M_\odot$, and $1/18 < q \leqslant 1$, respectively. 
We consider two prior choices for the aligned spins:
\begin{enumerate}
    \item {\it Isotropic spin prior:} For either binary component, the (dimensionless) spin vector $\bfchi$ is isotropically oriented, while the spin magnitude $|\bfchi|$ is drawn from a flat distribution within $[0,\,\chi_{\rm max}]$. We extract the aligned component $\chi_z$ and pass it to the waveform model.
    
    \item {\it Flat $\chi_{\rm eff}$ prior:} Both the aligned spins $\chi_{1z}$ and $\chi_{2z}$ are allowed to be in the range $[-\chi_{\rm max},\,\chi_{\rm max}]$. Given the values of the component masses, the joint prior for the aligned spin components is
    \begin{multline}
       P(\chi_{1z}, \chi_{2z})\,\rmd\chi_{1z}\,\rmd\chi_{2z} \propto \rmd\chi_{1z}\,\rmd\chi_{2z} \\
       \times 
       \begin{cases}
           1, & \quad |\chi_{\rm eff}| \leqslant \chi_{\rm max}\,\frac{m_1 - m_2}{M}, \\ 
           \frac{1-(m_1 - m_2)/M}{1 - |\chi_{\rm eff}|/\chi_{\rm max}}, & \quad |\chi_{\rm eff}| > \chi_{\rm max}\,\frac{m_1 - m_2}{M}.
       \end{cases} 
    \end{multline}
    where $M=m_1 + m_2$ is the total mass and the effective aligned-spin parameter is given by $\chi_{\rm eff} = (m_1\,\chi_{1z} + m_2\,\chi_{2z})/M$. 
    This prior is designed such that $\chi_{\rm eff}$ is distributed uniformly within $[-\chi_{\rm max},\,\chi_{\rm max}]$. 
\end{enumerate}
We choose $\chi_{\rm max} = 0.99$ in order to allow highly spinning mergers. 

To leading order, $\chi_{\rm eff}$ is the only spin parameter that determines the phasing of the gravitational waveform. 
The isotropic spin prior strongly penalizes configurations in which the two black holes have large and aligned spins, and hence the induced prior on $\chi_{\rm eff}$ is significantly peaked around $\chi_{\rm eff} = 0$. 
The isotropic prior is a natural consequence of dynamical models for BBH formation. 
In order to examine the impact of this assumption on parameter estimation, we contrast the results with those obtained using the flat prior on $\chi_{\rm eff}$ (as a proxy for other astrophysical scenarios).

\begin{figure}
    \centering
    \includegraphics[scale=0.36]{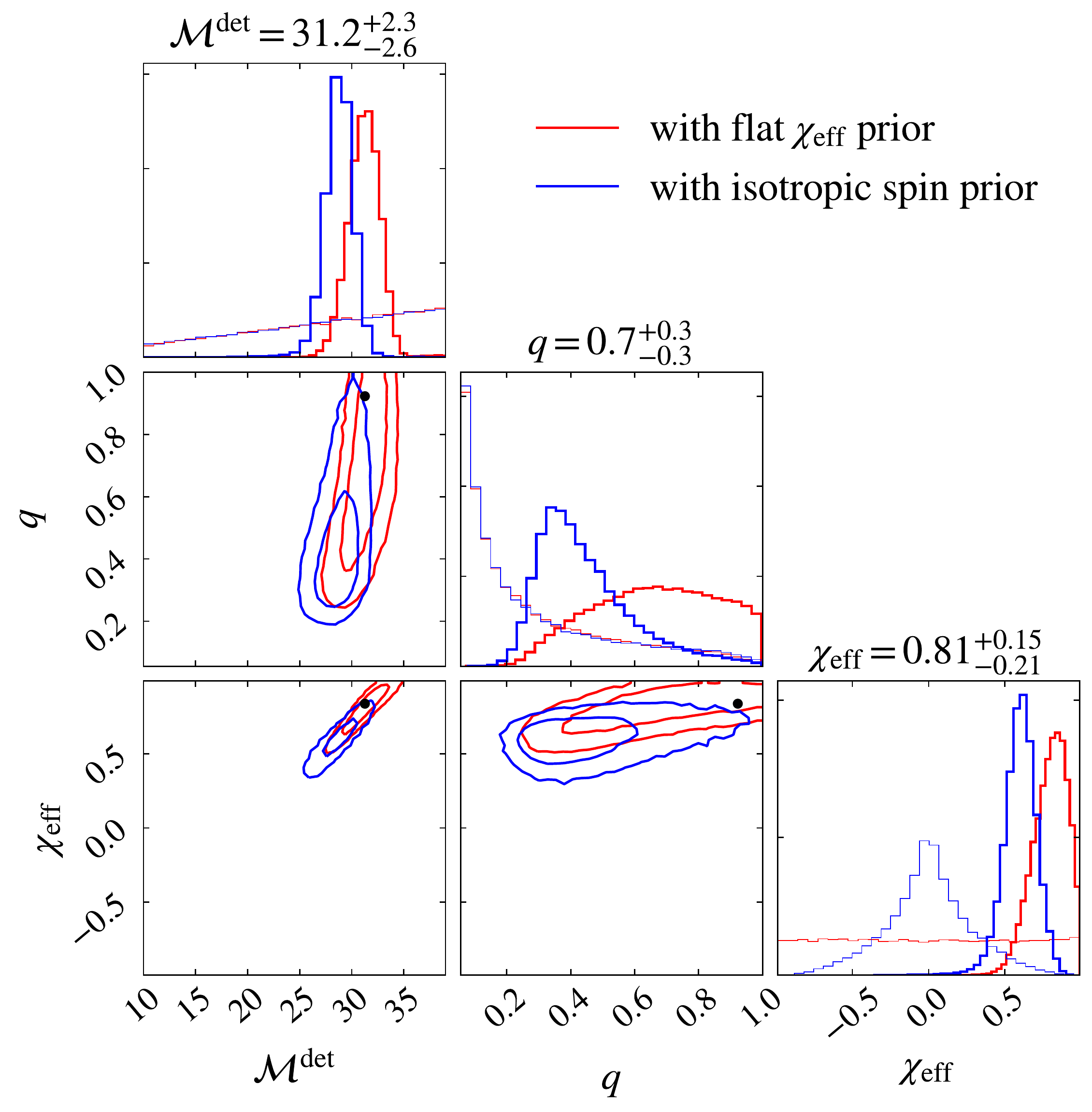}
    \caption{\label{fig:alignspin_post} Posterior distributions for the detector-frame chirp mass $\mathcal{M}^{\rm det}$, the mass ratio $q=m_2/m_1$, and the effective aligned spin $\chi_{\rm eff}$ obtained using the \texttt{IMRPhenomD} waveform model. We compare results obtained using the flat $\chi_{\rm eff}$ prior (red) and the isotropic spin prior (blue). The contours in the off-diagonal panels enclose 68\% and 95\% quantiles for the joint posterior distributions for each pair of parameters, and black dots mark the maximum-likelihood solution. The diagonal panels show the marginalized posterior (thick curves) and prior (thin curves) distributions. The parameter values quoted are the median and the 90\% credible uncertainty intervals obtained using the flat $\chi_{\rm eff}$ prior.}
\end{figure}

\reffig{alignspin_post} shows the posterior distributions for $\mathcal{M}^{\rm det}$, $q$ and $\chi_{\rm eff}$ under the two different spin priors. 
Under the isotropic spin prior, the most probable value for $\chi_{\rm eff}$ is $\simeq 0.55$, which is already higher than the values observed in previous LIGO/Virgo BBH events. 
In addition, the mass-ratio $q \sim 0.4$ is peaked away from unity. 
However, the isotropic spin prior peaks at $\chi_{\rm eff} = 0$, and is suppressed at $\chi_{\rm eff} \simeq 0.55$. 
This suggests that even higher values are penalized by the prior rather than by the data itself. 
Under the flat $\chi_{\rm eff}$ prior, we indeed measure a higher value for the aligned spin $\chi_{\rm eff} = 0.81^{+0.15}_{-0.21}$, which requires both black holes to be rapidly spinning and aligned. 
The mass ratio $q$ is also consistent with unity, and hence, the choice of spin prior also affects the most probable value for the chirp mass. 

The maximum likelihood solution has a strikingly high value of $\chi_{\rm eff}=0.84$ for the aligned spin, and a mass ratio $q \approx 1$. 
In all the two-dimensional marginalized joint posterior distributions of \reffig{alignspin_post}, the maximum likelihood parameters coincide with the most probable {\it a posteriori} values for the flat $\chi_{\rm eff}$ prior, but fall nearly outside the $95$\% quantiles for the isotropic prior. 
More formally, the Bayesian evidence ratio between the flat $\chi_{\rm eff}$ prior and the isotropic spin prior is $K \approx 22$, which favors the former prior choice over the latter in the sense of Bayesian model selection~\cite{jeffreys1998theory}. 

\reftab{parameters} summarizes the source parameters and their uncertainties under the two different spin priors. The results demonstrate the impact of the choice of priors on GW parameter estimation~\cite{Vitale:2017cfs}. 
Astrophysical mechanisms that involve binary stellar evolution can form aligned and highly spinning black hole binaries~\cite{zaldarriaga2017expected}; thus, we should take care to adopt priors that do not unfairly penalize solutions with large aligned (or anti-aligned) spins.

\begin{table}[]
\begin{center}
\setlength\tabcolsep{3pt}
\begin{tabular}{L{3.85cm}C{1.8cm}C{2cm}}
\specialrule{.1em}{.05em}{.05em} 
\specialrule{.1em}{.05em}{.05em} 
& Flat $\chi_{\rm eff}$ prior & Isotropic spin prior \\ 
\specialrule{.1em}{.05em}{.05em} 
Chirp mass $\mathcal{M}^{\rm det}$ & $31^{+2}_{-3}\,M_\odot$ & $29^{+2}_{-2}\,M_\odot$ \\
Primary mass $m_1$       & $31^{+13}_{-6}\,M_\odot$ & $38^{+11}_{-11}\,M_\odot$ \\
Secondary mass $m_2$     & $21^{+5}_{-6}\, M_\odot$ & $16^{+6}_{-3}\,M_\odot$ \\
Mass ratio $m_2/m_1$                & $0.7^{+0.3}_{-0.3}$  & $0.4^{+0.3}_{-0.1}$ \\
Total mass $M$           & $52^{+9}_{-6}\, M_\odot$ & $54^{+10}_{-8}\,M_\odot$ \\
Primary aligned spin $\chi_{1z}$ & $0.86^{+0.12}_{-0.27}$ & $0.73^{+0.18}_{-0.28}$   \\
Secondary aligned spin $\chi_{2z}$ & $0.79^{+0.19}_{-0.65}$ & $0.30^{+0.51}_{-0.46}$   \\
Effective aligned spin $\chi_{\rm eff}$ & $0.81^{+0.15}_{-0.21}$ & $0.60^{+0.16}_{-0.18}$   \\
Cosine of inclination $|\cos\iota|$ & $0.81^{+0.18}_{-0.52}$  & $0.81^{+0.18}_{-0.51}$ \\
Luminosity distance $D_L$ & $2.4^{+1.2}_{-1.1}\, {\rm Gpc}$ & $2.1^{+1.0}_{-0.9}\, {\rm Gpc}$ \\
Source redshift $z$                 & $0.43^{+0.17}_{-0.17}$ & $0.38^{+0.15}_{-0.15}$ \\
\specialrule{.1em}{.05em}{.05em} 
\specialrule{.1em}{.05em}{.05em} 
\end{tabular}
\caption{\label{tab:parameters} Source properties for \CANDIDATE: we give uncertainties encompassing the $90\%$ credible intervals in the posterior distribution under two different assumptions about the prior distribution of black hole spins. Parameter estimations were performed with the spin-aligned waveform model \texttt{IMRPhenomD}. All masses are quoted in the source frame except that the chirp mass $\mathcal{M}^{\rm det}$ is quoted in the detector frame. }
\end{center}
\end{table}

{\bf Spin-misaligned model}. In this section, we expand the parameter space to allow for misaligned component spins, and examine the gravitational wave data for evidence for the associated spin-orbit precession. To date spin-orbit precession has not been detected in any of the reported BBH merger events.

We use the waveform model \texttt{IMRPhenomPv2}, which phenomenologically models the waveform in the presence of spin-orbit precession~\cite{Schmidt:2014iyl, Hannam:2013oca}. 
We adopt the same priors for the masses and for the extrinsic parameters, except that we do not include the luminosity distance $D_L$ as an explicit parameter, but rather maximize the likelihood with respect to the common normalization of the signals to reduce the computational cost.

We use a spin prior that is similar to the flat $\chi_{\rm eff}$ prior in the case of the spin-aligned analysis.
In this case, we sample spin vectors with all orientations and with the spin magnitudes within the range $[0,\,\chi_{\rm max}]$. 
For given component masses, we assign a joint prior:
\begin{multline}
    \label{eq:generic_spin_prior}
    P\left(\bfchi_1,\,\bfchi_2\right)\,\rmd^3\bfchi_1\,\rmd^3\bfchi_2 \propto  \frac{\rmd^3\bfchi_1}{\chi^2_{\rm max} - \chi^2_{1z}}\,\frac{\rmd^3\bfchi_2}{\chi^2_{\rm max} - \chi^2_{2z}} \\
    \times
    \begin{cases}
    1,& \quad |\chi_{\rm eff}| \leqslant \chi_{\rm max}\,\frac{m_1 - m_2}{M}, \\ 
    \frac{1-(m_1 - m_2)/M}{1 - |\chi_{\rm eff}|/\chi_{\rm max}}, & \quad |\chi_{\rm eff}| > \chi_{\rm max}\,\frac{m_1 - m_2}{M}.
    \end{cases}
\end{multline}
With all Cartesian spin components marginalized over, $\chi_{\rm eff}$ has a uniform distribution in the range $[-\chi_{\rm max},\,\chi_{\rm max}]$.
Again, we set $\chi_{\rm max}=0.99$ in our analysis.

Our strategy is to perform a fully precessing analysis with \texttt{IMRPhenomPv2}, and in addition design a control test:
\begin{enumerate}
    \item {\it Precession test:} Generate spin vectors using Eq.\eqref{eq:generic_spin_prior} and pass all Cartesian components to \texttt{IMRPhenomPv2}.
    \item {\it Control test:} Generate spin vectors using Eq.\eqref{eq:generic_spin_prior}. Pass only the aligned components to \texttt{IMRPhenomPv2}, but pass zeros for the in-plane components. 
\end{enumerate}
If waveforms with precession genuinely fit the data better than non-precessing waveforms, the precession and control tests should yield different results.

The leading effect of spin-orbit precession can be captured by a single parameter $\chi_p$, which is defined to be~\cite{Schmidt:2014iyl}
\be
\chi_p \coloneqq \frac{1}{A_1\,m^2_1} \max_{}\left( A_1\,\left|\bfchi_{1,\perp}\right|\,m_1^2,\,A_2\,\left|\bfchi_{2,\perp}\right|\,m_2^2 \right).
\ee
where $A_1 = 2+ (3\,q)/2$ and $A_2 = 2 + 3/(2\,q)$, and $\bfchi_{1,\perp}$ and $\bfchi_{2,\perp}$ are the spin vectors perpendicular to the orbital plane, for the primary and the secondary respectively. 

\begin{figure}
    \centering
    \hspace{-0.5cm}\includegraphics[scale=0.5]{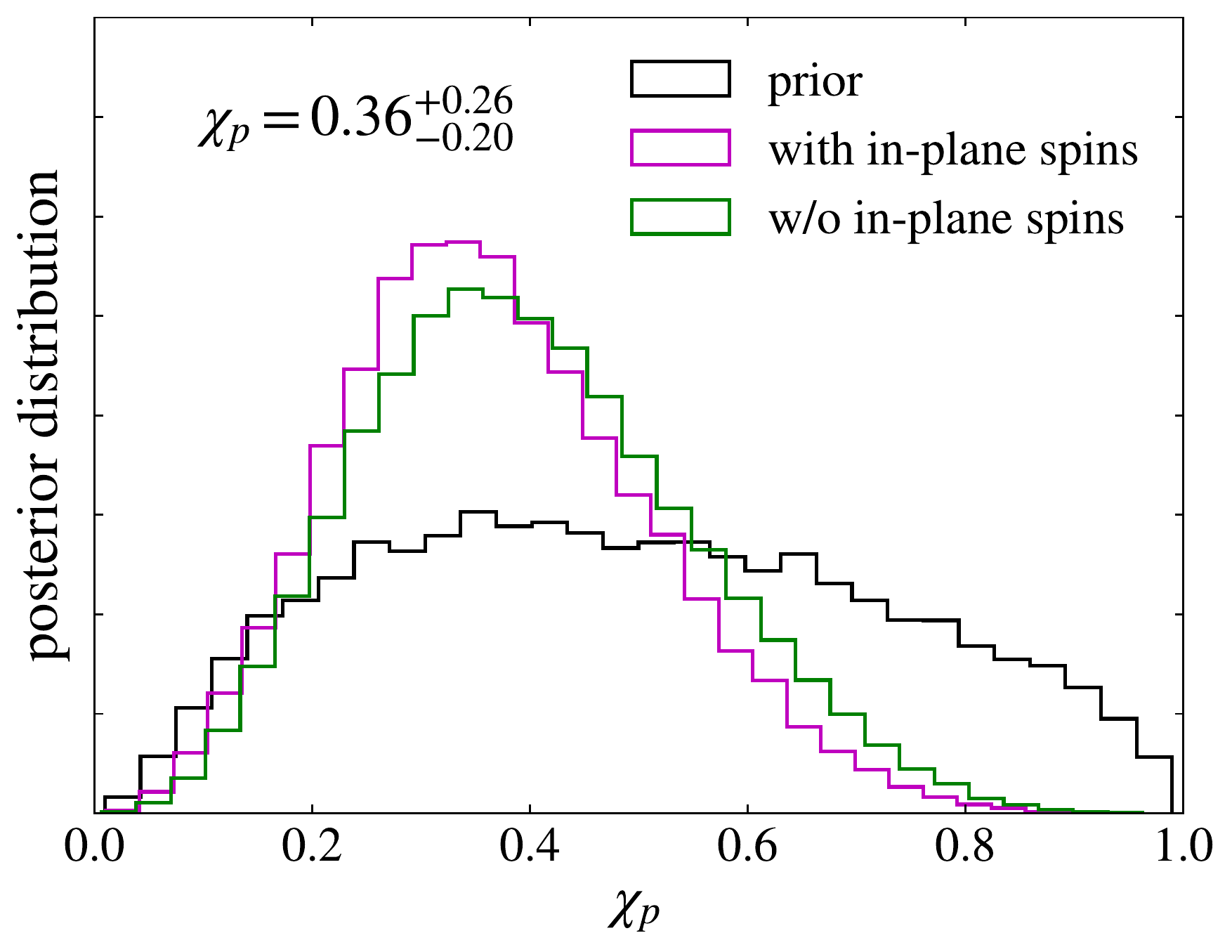}
    \caption{\label{fig:chip_post_Pv2} Prior and posterior distributions for the effective spin-precession parameter $\chi_p$ obtained using the \texttt{IMRPhenomPv2} waveform model. We compare the results obtained using all the spin parameters (magenta) and by passing zero in-plane spin components to the waveform generation routine, without changing the rest of the prior (green). The consistency of these two curves illustrates that the detected signal has no signs of precession. The value and range for $\chi_p$ are the median and the 90\% credible uncertainty range, respectively.}
\end{figure}

\begin{figure*}
    \centering
    \hspace{-0.5cm}\includegraphics[scale=1.0]{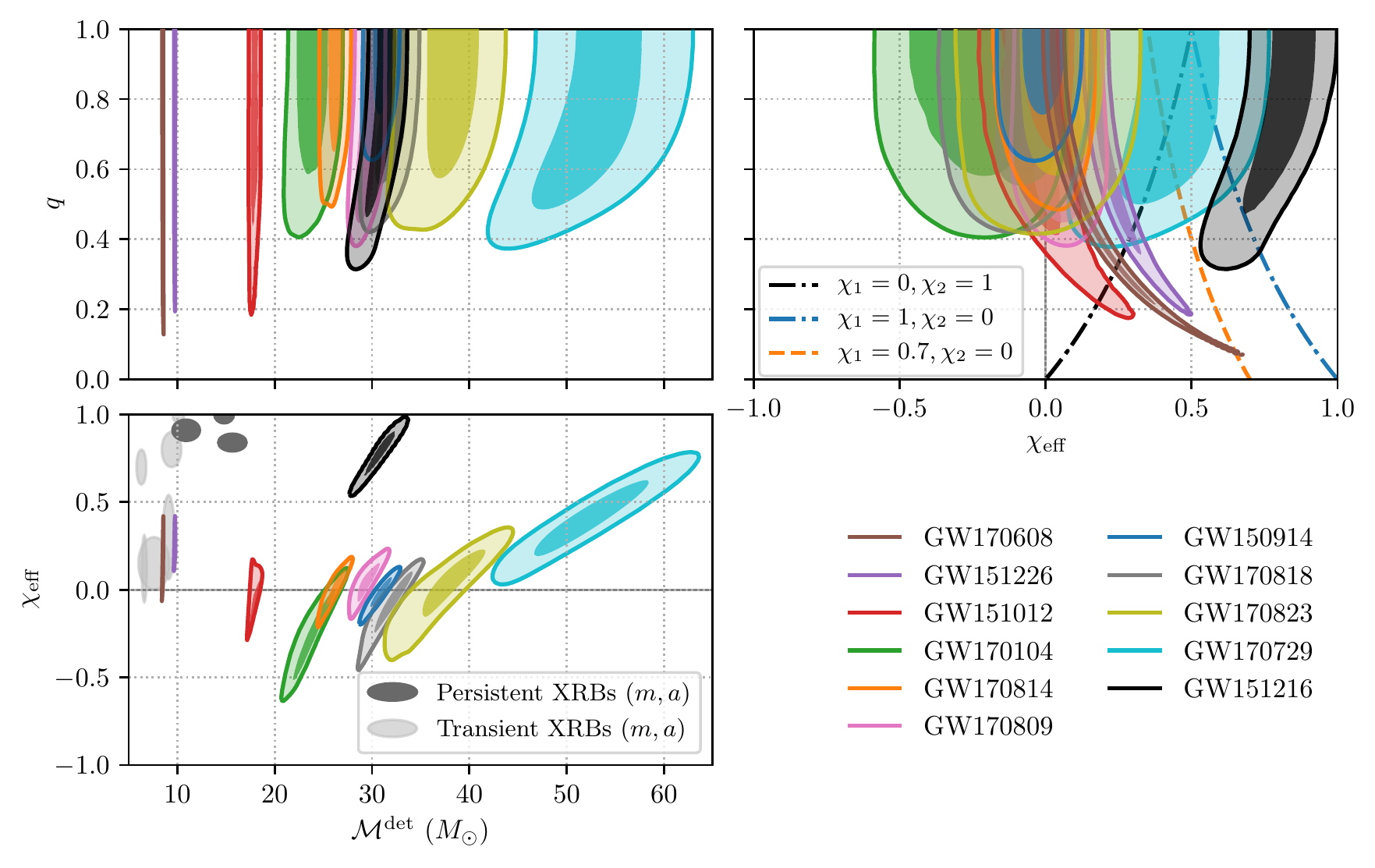}
    \caption{\label{fig:bbh_dist} Marginalized likelihood contours enclosing $50\%$ and $90\%$ of the distribution for BBH mergers detected to date. Likelihoods are computed using the frequency-domain surrogate model \texttt{SEOBNRv4\_ROM}~\cite{Bohe:2016gbl}, which is in good agreement with the analysis using \texttt{IMRPhenomD}. The panel in the lower left contrasts the populations of the detected mergers, and persistent and transient X-ray binaries reported in Ref.~\cite{mcclintock2013black}, in the $(\mathcal{M}, \chi_{\rm eff})$ plane. Dashed-dotted lines in the right-hand panel mark the allowed parameter space when the aligned spins of the black holes take specific values, which are typical of scenarios in which the BBH progenitors are tidally locked, or formed dynamically.}
\end{figure*}

\reffig{chip_post_Pv2} shows the posterior distributions for $\chi_p$ for the precession and control tests. 
In both cases, the posteriors for the masses, aligned spin components, and the extrinsic parameters are consistent with those in the aligned-spin-only case (with the uniform $\chi_{\rm eff}$ prior).
The posterior for $\chi_p$ appears significantly narrower than the prior distribution, but the control test yields nearly identical results. 
This suggests that there is no direct information about spin-orbit precession, instead, the data tightly constrains $\chi_{\rm eff}$, which then restricts the allowed values of $\chi_p$ due to physical constraints on the spins. 
The maximum likelihood improves by about one unit for \texttt{IMRPhenomPv2} compared to \texttt{IMRPhenomD}; this is however not significant due to the larger number of free parameters in the former. 


A prior analysis of a different GW event (GW151226) that measured a non-zero value for $\chi_{\rm eff} \approx 0.2$ also reported a posterior distribution for $\chi_p$ that differed noticeably from the prior~\cite{Abbott:2016nmj}. 
For the reasons mentioned above, we should exercise caution in interpreting these posteriors as evidence for precession in the data.

\section{Possible formation channels}

A number of formation channels for binary black hole mergers have previously been suggested in the literature, including isolated binaries of massive stars that evolve through a common envelope phase~\cite{nelemans2001gravitational, belczynski2002comprehensive, voss2003galactic, belczynski2007rarity, belczynski2008compact, dominik2013double, belczynski2014formation, mennekens2014massive, spera2015mass, eldridge2016bpass, stevenson2017formation, mapelli2017cosmic, giacobbo2017merging, mapelli2018cosmic, kruckow2018progenitors, giacobbo2018progenitors}, or through a phase of chemically homogeneous evolution~\cite{marchant2016new, de2016chemically, mandel2016merging}, few-body interactions at the core of dense stellar environments, such as old globular clusters~\cite{zwart1999black, o2006binary, sadowski2008total, downing2010compact, downing2011compact, Samsing:2013kua, PhysRevLett.115.051101, rodriguez2016binary, askar2016mocca}, young open clusters~\cite{ziosi2014dynamics, mapelli2016massive, banerjee2017stellar, chatterjee2017dynamical}, or nuclear clusters at the center of galaxies~\cite{antonini2016merging, petrovich2017greatly}. 
It has also been suggested that binaries are driven toward merger by nearby supermassive black holes~\cite{antonini2012secular}, their accretion disks~\cite{mckernan2012intermediate, stone2016assisted, bartos2017rapid}, or by tertiary stellar companions~\cite{antonini2014black, kimpson2016hierarchical, antonini2017binary, liu2018black}. 
Finally, it has been suggested that BBHs are remnants of Population III stars~\cite{Kinugawa:2014zha, inayoshi2016gravitational}, or relics of the primordial universe~\cite{carr1974black, carr2016primordial, sasaki2016primordial, inomata2017inflationary, bird2016did, blinnikov2016solving, ali2017merger, clesse2017clustering, chen2018merger, ando2018primordial}. 
Measurements of the masses and spins are crucial to distinguishing between these formation scenarios~\cite{Farr:2017gtv}.

\reffig{bbh_dist} compares the parameters inferred for \CANDIDATE (using the flat $\chi_{\rm eff}$ prior) to those of the other O1/O2 BBH events. We see that the component masses are similar to those of the heavy O1/O2 BBH events, but are higher than those of the known high-mass X-ray binaries~\cite{Bogomazov:2016cei, corral2016blackcat}. 
The inferred mass ratio is close to unity, which is consistent with both isolated binary evolution~\cite{belczynski2014formation, Kinugawa:2014zha, eldridge2016bpass, stevenson2017formation, marchant2017ultra} and dynamical formation~\cite{rodriguez2016binary, Park:2017zgj}.

The effective aligned-spin parameter $\chi_{\rm eff}$, inferred using the flat $\chi_{\rm eff}$ prior, is higher than that of any of the O1/O2 BBH events; there is even substantial probability for $\chi_{\rm eff} > 0.9$. 
For a mass-ratio close to unity, this requires that both black holes are nearly maximally spinning along the direction of the orbital angular momentum, which is improbable if the binary formed dynamically. 
The most probable spin value of $\chi_{\rm eff} \simeq 0.6$ under the isotropic prior is easier to reconcile with this scenario. 
However, the mass ratio is driven to $q \sim 1/2$--$1/3$ in this case. 
This is atypical of dynamical formation scenarios, in which binaries harden through successive binary-single interactions. 
Mergers with large mass-ratios are rare, and can occur if the heavier primary itself is the outcome of a previous BBH coalescence that was retained by the stellar cluster~\cite{Samsing:2017jnz, Rodriguez:2017pec}. 
If the seed black holes are non-spinning, the merger product (our putative primary) typically has a spin of $\chi_{1z} \simeq 0.7$; the likelihood of \CANDIDATE nearly rules out the case in which the primary has $\chi_{1z} \simeq 0.7$ and the secondary has a low spin (right panel in \reffig{bbh_dist}). 
Joint consideration of spin and mass ratio therefore disfavors the possibility that \CANDIDATE was dynamically formed. 

Stellar binary evolution naturally leads to aligned mergers, with the caveat that the spin(s) and the orbit can be misaligned for sufficiently large BH natal kicks~\cite{o2017inferences, wysocki2018explaining}. 
In particular, remnants can have large aligned spins after the progenitor stars are tidally locked, even if the progenitor stars are slowly spinning initially~\cite{kushnir2016gw150914,zaldarriaga2017expected, Hotokezaka:2017dun, hotokezaka2017implications}. The separation required for a binary to become tidally locked is comparable to that required for it to merge as a result of gravitational radiation in the age of the Universe; thus one might expect that the BH formed last would have a high spin in a fair fraction of BBH mergers~\cite{zaldarriaga2017expected}. \reffig{bbh_dist} shows that this picture is inconsistent with the data from this event if the lightest of the two black holes formed last, but consistent if the heavier one is the last to form. Alternatively, if the binary became tight and tidally locked before either BH had formed, both BHs would be spinning rapidly in good agreement with the parameters derived for this system.

Tidally locked binaries have smaller separations before the second star collapses into a black hole, and thus have shorter delay times between binary formation and merger. \CANDIDATE has the second highest redshift $z \simeq 0.43$ among the population of detections, only below GW170729 (another massive BBH merger with $z \simeq 0.48$, which incidentally has the second highest spin-parameter $\chi_{\rm eff} \simeq 0.36$~\cite{LIGOScientific:2018mvr}). Intriguingly, Ref.~\cite{hotokezaka2017implications} argued that within the tidal-locking scenario, fast spinning BBHs are more prevalent at higher redshifts $(z \sim 0.5$ to $1.5)$ than in the present-day universe.

\section{Conclusions}

We report a new BBH merger in the public data from the First Observing Run of advanced LIGO. 
The candidate has a detector-frame chirp mass $\mathcal{M}^{\rm det} = 31^{+2}_{-3}\,M_\odot$, and has a FAR of 1 per 52 O1 in the bank with chirp-masses $\mathcal{M}^{\rm det} \in 20 - 40 \, M_\odot$.
It is among the highest redshift events discovered thus far. 
The inferred value for the effective spin parameter $\chi_{\rm eff} \simeq 0.8$ is the highest among the BBHs detected to date, and points to the system consisting of two rapidly spinning black holes, with spins aligned with the orbital plane. 
Future detections of BBH mergers similar to our candidate can confirm the existence of a population of fast-spinning and aligned mergers; correlations between the spins, mass-ratios, and source redshifts can shed light on the astrophysical origin of these mergers. Such a population is unlikely in dynamical formation scenarios, but is characteristic of stellar binary evolution. 

\section*{Acknowledgment}

We thank the participants of the JSI-GWPAW 2018 Workshop held at the University of Maryland and the GWPOP conference held in Aspen (2019) for the constructive discussions and comments.
We thank David Radice for illuminating discussions.
We greatly thank the LIGO Collaboration and the Virgo Collaboration for making the O1 data publicly accessible and easily usable.

This research has made use of data, software and/or web tools obtained from the Gravitational Wave Open Science Center (https://www.gw-openscience.org), a service of LIGO Laboratory, the LIGO Scientific Collaboration and the Virgo Collaboration. LIGO is funded by the U.S. National Science Foundation. Virgo is funded by the French Centre National de Recherche Scientifique (CNRS), the Italian Istituto Nazionale della Fisica Nucleare (INFN) and the Dutch Nikhef, with contributions by Polish and Hungarian institutes.

BZ acknowledges the support of The Peter Svennilson Membership
fund. LD acknowledges the support from the Raymond and Beverly Sackler Foundation Fund. TV acknowledges support by the Friends of the Institute for Advanced Study.
MZ is supported by NSF grants AST-1409709,  PHY-1521097 and  PHY-1820775 the Canadian Institute for Advanced Research (CIFAR)
program on Gravity and the Extreme Universe and the Simons Foundation Modern Inflationary Cosmology initiative.




\bibliographystyle{apsrev4-1-etal}
\bibliography{bbh}

\end{document}